\documentclass[twocolumn, showpacs,preprintnumbers,amsmath,amssymb]{revtex4}
\usepackage{enumitem}
\usepackage{amssymb}
\usepackage{xcolor}
\usepackage{graphicx}% Include figure files
\usepackage{dcolumn}% Align table columns on decimal point
\usepackage{bm}% bold math
\usepackage{multirow}
\usepackage{booktabs}
\usepackage{longtable}
\usepackage{natbib}
\usepackage{lscape}
\usepackage{soul}
\usepackage{amssymb}
\usepackage{amsmath}	% Advanced maths commands
\usepackage{color}
\usepackage{cases} 

\usepackage{array}

\begin{document}
\preprint{APS/123-QED}

\title{Low-energy collisions between electrons and BeD$^+$}
\author{S. Niyonzima$^{1,2}$}
\author{N. Pop$^{3}$}
\author{F. Iacob$^{4}$}
\author{ \AA. Larson$^{5}$}
\author{A.E. Orel$^{6}$}
\author{J. Zs Mezei$^{2,7,8}$}%\email[]{mezei.zsolt@atomki.hu}
\author{K. Chakrabarti$^{9}$}
\author{V. Laporta$^{2,10}$}
\author{K. Hassouni$^{7}$}
\author{D. Benredjem$^{11}$}
\author{A. Bultel$^{12}$}
\author{J. Tennyson$^{10}$}
\author{D. Reiter$^{13}$}
\author{I. F. Schneider$^{2,11}$}\email[]{ioan.schneider@univ-lehavre.fr}
\affiliation{$^{1}$D\'ept. de Physique, Facult\'e des Sciences, Universit\'e du Burundi, B.P. 2700 Bujumbura, Burundi}%
\affiliation{$^{2}$LOMC CNRS$-$Universit{\'{e}} du Havre$-$Normandie Universit{\'{e}}, 76058 Le Havre, France}
\affiliation{$^{3}$Dept. of Fundamental of Physics for Engineers, Politehnica University Timisoara,  300223 Timisoara, Romania}%
\affiliation{$^{4}$Dept. of Physics, West University of Timișoara, 300223 Timisoara, Romania}
\affiliation{$^{5}$Dept. of Physics, Stockholm University, AlbaNova University Center, 106 91 Stockholm, Sweden}%
\affiliation{$^{6}$Dept. of Chemical Engineering, University of California, Davis, California 95616, USA}%
\affiliation{$^{7}$LSPM, CNRS$-$Universit\'e Paris 13$-$USPC, 93430 Villetaneuse, France}%
\affiliation{$^{8}$HUN-REN Institute for Nuclear Research (ATOMKI), H-4001 Debrecen, Hungary}%
\affiliation{$^{9}$Dept. of Mathematics, Scottish Church College, Calcutta 700 006, India}
\affiliation{$^{10}$Dept. of Physics and Astronomy, University College London, WC1E 6BT London, UK}%
\affiliation{$^{11}$LAC, CNRS$-$Universit\'e Paris-Sud$-$ENS Cachan$-$Universit\'e Paris-Saclay, 91405 Orsay, France}%
\affiliation{$^{12}$CORIA CNRS$-$Universit\'{e} de Rouen$-$Universit{\'{e}} Normandie, F-76801 Saint-Etienne du Rouvray, France}%
\affiliation{$^{13}$IEK, Forschungszentrum J\"ulich GmbH Association EURATOM-FZJ, Partner in Trilateral Cluster, 52425 J\"ulich, Germany}%
\date{\today}

\begin{abstract}
Multichannel quantum defect theory is applied in the treatment of the
dissociative recombination and vibrational excitation processes for the BeD$^+$ ion
in the twenty four vibrational levels of its ground electronic state
($\textrm{X}\,{^{1}\Sigma^{+}},v_{i}^{+}=0\ldots 23$). Three electronic
symmetries of BeD$^{**}$ states (\ensuremath{^{2}\Pi},
\ensuremath{^{2}\Sigma^{+}}, and \ensuremath{^{2}\Delta}), are considered in
the calculation of cross sections and the corresponding rate coefficients.
The incident electron energy range is $10^{-5}$--2.7 eV and the electron temperature range is 100--5000~K. The
vibrational dependence of these collisional processes is highlighted. The
resulting data are useful in magnetic confinement fusion edge plasma modelling
and spectroscopy, in devices with beryllium based main chamber materials, such
as ITER and JET, and operating with the deuterium-tritium fuel mix. An
extensive rate coefficients database is presented in graphical form and also by
 analytic fit functions whose parameters are tabulated
in the supplementary material.
\end{abstract}

%\keywords{coupled-channel, optical shielding, KCs}%Use showkeys class option if keyword

\maketitle

\section{Introduction}

The  International Thermonuclear Experimental Reactor (ITER) is
aimed at demonstrating the scientific  and  technological feasibility  of fusion
power~\cite{Kleyn2006}. It is now widely accepted by the fusion community that
some form of controlled thermonuclear reactor, capable of producing a useful
amount of electrical power, will be built-in the not-too-distant future. To
obtain tenfold power multiplication in a controlled fusion process, at a power
level greater than 500 MW and during pulses of 10 min or longer,
exothermic reactions involving light nuclei, those between the hydrogen
isotopes, are by far the most probable and efficient. The Joint European Torus
(JET), in operation since 1983, has been persistently upgraded, most recently to become an
ITER-like wall. It's the largest and most powerful tokamak in the world capable
of operating with the deuterium-tritium fuel mix. One of the main improvements
of JET was to equip the vessel with a first wall material combination comprising
beryllium (Be) in the main chamber and tungsten (W) in the
divertor~\cite{Duxbury1998, Brezinsek2014, Brezinsek2015}. These
plasma facing components are expected to improve the machine conditioning, impact
on operational space and energy confinement. The installation of Be and W in the
main chamber wall of JET is aimed at studying the impurity evolution and
material migration under plasma and material conditions relevant for
ITER~\cite{Coenen2013}. The selection of beryllium in the main chamber wall is
explained by its operational flexibility anticipated for a low-Z main
wall~\cite{Matthews2013}, its low-fuel retention and excellent oxygen getter property,
confirmed experimentally~\cite{Brezinsek2015, Brezinsek2013}. In tokamaks, material erosion
limits the lifetime of plasma-facing components, while in the edge and divertor
regions of fusion reactors, plasma-wall interactions generate new molecular
species, this formation of impurities being allowed by the relatively
low-temperatures of this region of the fusion plasma. Moreover, due to the
strong chemical affinity of beryllium and tungsten to oxygen, those surfaces
will be oxidized~\cite{Alimov2004}.

In JET, the formation of BeD as well as the presence of Be, Be$^+$, BeD$^+$, BeT$^+$ and other
impurities into the plasma are clearly described in~\cite{Duxbury1998,niyonzima2016}
and experimentally confirmed by spectroscopic methods~\cite{Duxbury1998, Brezinsek2014,
Brezinsek2015, Krieger2013, Nishijima2008, Doerner2009}. Be erosion as well as its continuous
deposition towards the divertor is intrinsic to plasma operation due to the relatively high
 chemically assisted physical sputtering yield of Be \emph{via} the radical BeD
 molecule~\cite{Brezinsek2015} which dissociates by the reactions~\cite{Brezinsek2014}:
\begin{equation}
e + \textrm{BeD} \to \textrm{Be} + \textrm{D} + e\,.
\end{equation}
Though deuterium ion bombardment of Be targets may cause the formation of the
stable BeD$_2$ molecule near the main chamber, there is no spectroscopic access
to the BeD$_2$ molecule  released by chemically assisted physical sputtering of
Be wall~\cite{Brezinsek2014, Nishijima2008}. Furthermore, retention of fuel
elements by implantation in Be is expected to be saturated quickly due to the
narrow interaction zone~\cite{Brezinsek2013}. With the full W divertor installed
in JET, all Be ions flowing into the inner divertor~\cite{Krieger2013} are
originated primarily in the main chamber during diverted plasma operation.
Finally, the main physics mechanism responsible for the fuel retention under the
Be wall conditions in the JET experiments is co-deposition of fuel in Be
co-deposits~\cite{Brezinsek2013}. The rate of fuel retention with the ion flux
to the main plasma facing components in both the divertor and main chamber is
increased by co-deposition of fuel atoms with Be~\cite{Matthews2013,
Doerner2009}. This information is in line with the measured spectral line
emission of BeII (Be$^+$) influx~\cite{Brezinsek2015, Nishijima2008} from the
main chamber into the inner divertor whose plasma-facing surfaces are a net
deposition zone~\cite{Krieger2013}.

In tokamaks with Be/W environment, all studies leading to physics understanding of
beryllium migration and connecting the lifetime of the first wall components under
erosion, with tokamak safety, in relation to the temporal behaviour of each
fraction contribution to the long-term retention. The
erosion mechanism itself is not studied in this work, but we are interested in providing
data to support diagnosing beryllium in the  fusion plasma. Moreover, in JET
equiped with Be/W wall environment and operating with deuterium-tritium fuel
mix, the rate of Be erosion is measured by spectroscopy of all the states of the
atoms and molecules, so primarily of Be, Be$^+$, Be$_2^+$, BeD, BeD$^+$,
BeT, BeT$^+$, BeD$_2$, BeDT, BeT$_2$, BeD$_2^+$, BeDT$^+$ and BeT$_2^+$. Several observations of
Be erosion by optical emission spectroscopy of various transitions of Be (at 457
nm)~\cite{Duxbury1998,Brezinsek2014}, Be$^+$ (at 527 nm and
436nm)~\cite{Brezinsek2014} and the $\textrm{A}\,{^{2}}\Sigma{^+} \to
\textrm{X}\,{^{2}}\Sigma{^+}$ band emission of BeD (band head at 497-500
nm)~\cite{Nishijima2008,jtBeHDTfit} under different plasma conditions and surface
temperatures, have been carried out successfully in laboratories and JET
experiments.

As shown in Ref~\cite{Duxbury1998}, BeD and BeD$^+$ are the only beryllium hydride molecules
released in the plasma. Even though BeD$^+$ is expected to be stable in the JET divertor
plasma~\cite{Coxon1997}, being formed through the reactions~\cite{Brezinsek2014,
Nishijima2008},
\begin{eqnarray}
e + \textrm{BeD} &\to& \textrm{BeD}^+ + e + e\,,\\
\textrm{Be}^+ + \textrm{D}_2 &\to& \textrm{BeD}^+ + \textrm{D}\,,
\end{eqnarray}
its $\textrm{A}\,{^{1}\Sigma{^+}} \to \textrm{X}\,{^{1}\Sigma}{^+}$ band
emission in visible and ultra-violet range is not observable, probably due to its
weak intensities~\cite{Coxon1997, Koontz1935}. Be is the main and dominant intrinsic impurity in limited and
diverted plasmas with the JET. Those impurities hugely influence the low
temperature edge and divertor plasma behaviour in which electrons and ions
originating from the core plasma are cooled by radiation and charge exchange
processes till below 1 eV~\cite{Kleyn2006, McCracken1998, Krasheninnikov2002}.
All molecular species in these regions undergo many collisions in particular
those between electrons and molecular ions are of crucial importance~\cite{McCracken1998}. The
electron-impact processes of vibrationally excited BeD$^+$ play a key role in
the reaction kinetics of low-temperature plasmas in general, and particularly
also in certain cold regions of fusion reactor relevant (\emph{e.g.} the
divertor) plasmas. Hence, modelling and diagnosing these varied plasma
environments require accurate, reliable cross-sections and rate coefficients for
interactions of these molecular ions with electrons~\cite{Celiberto2001, Reiter2012} which
produce simpler species, most of which being unsuitable for visible
spectroscopy. The present complete database of cross-sections and rate coefficients for
electron-impact collision processes coupled to the availability of absolutely
calibrated spectroscopic emission from this molecular ion provides a way to
characterise also the BeD$^+$ formation rates in the edge and divertor plasma of
fusion devices.

This work is a part of a series of papers~\cite{8b, niyonzima2016, 0741-3335-59-4-045008, niyonzima2013, jt692} devoted to the study of electron-impact processes in fusion devices with beryllium-based main chamber materials. This series was dedicated to the BeH$^+$ and BeH species, but they also contained preliminary studies of the isotopic effects. According to Figure 5 from \cite{8b} and Figure 11 from \cite{0741-3335-59-4-045008}, these effects are quite notable, and this has pushed us to address it in a systematic and exhaustive manner. In this article,  we present reactive collisions cross sections and rate coefficients between electrons and the BeD$^+$ molecular ion in all vibrationnal states, relevant for the divertor and edge plasma kinetics of JET and ITER. In collision with electrons the BeD$^{+}$ ion undergoes several processes, in particular dissociative recombination (DR) and vibrational-excitation/de-excitation (VE/VdE), respectively \cite{dr2013,roos:09}:
\begin{eqnarray}
e + \mathrm{BeD}^{+}(v_{i}^{+}) &\to& \mathrm{Be + D}\,,\hspace{1.5 cm} (\textrm{DR})\\
e + \mathrm{BeD}^{+}(v_{i}^{+}) &\to& \mathrm{BeD}^{+}(v_{f}^{+})+\mathrm{e}^{-}\,,\hspace{0.2 cm}  (\textrm{VE/VdE})
\end{eqnarray}
\noindent
where $v_{i}^{+}$($v_{f}^{+}$) denote the initial(final) vibrational level of the cation. At an energy exceeding the dissociation energy of BeH$^+$ (calculated to be $2.68$ eV, see below) also the process of dissociative excitation sets it. It is not considered in the present paper.

The manuscript is organized as follows: In Section~\ref{theoretical}, we briefly review the theoretical method used to calculate the cross sections and the corresponding rate coefficients; Section~\ref{res_disc} presents Maxwellian isotropic rate coefficients computed for the DR, VE and VdE processes. These rate coefficients have been fitted with a modified Arrhenius law. Section~\ref{sec:concl} contains the final remarks concluding the paper.

\section{Brief description of the theoretical approach of the dynamics \label{theoretical}}

In the present paper, we used the Multichannel Quantum Defect Theory
(MQDT)-type approach \cite{Giusti:80} to study the vibrational resolved reactive collisions of beryllium deuteride
cation (BeD$^{+}$) with electrons.
We assumed that BeD$^{+}$ is initially in its electronic ground state,
$\textrm{X}\,^1\Sigma^+$, with  energies below its dissociation limit.
The electron-impact collision processes covered by the present article involve two mechanisms which are treated simultaneously by the MQDT~\cite{Giusti:80}: (\textit{i}) the \textit{direct} process, in which the electron is captured into a doubly excited resonant state BeD$^{**}$ of the neutral system, resulting in two neutral atomic fragments Be and D or in autoionization,
\begin{equation}\label{eq:direct}
e + \mathrm{BeD}^{+}(v_{i}^{+})  \to \mathrm{BeD}^{**} \to  \left\{
			\begin{array}{ll}
			 \mathrm{Be + D}\\
			 \mathrm{BeD}^{+}(v_{f}^{+}) + e\,,\\
			  \end{array}
			  \right.
\end{equation}
and (\textit{ii}) the \textit{indirect} process consisting in the temporary capture of the electron into BeD$^{*}$, a singly excited bound Rydberg state, predissociated by BeD$^{**}$,
\begin{equation}\label{eq:indirect}
e + \mathrm{BeD}^{+}(v_{i}^{+})  \to \mathrm{BeD}^{*}\to \mathrm{BeD}^{**} \to  \left\{
			\begin{array}{ll}
			 \mathrm{Be + D}\\
			 \mathrm{BeD}^{+}(v_{f}^{+}) + e\,.\\
			  \end{array}
			  \right.
\end{equation}

In the MQDT approach, the processes (\textit{i}) and (\textit{ii}) result in the
total mechanism by quantum interference. As mentionned in Eqs.~(\ref{eq:direct})
and (\ref{eq:indirect}), the excited neutral system, reached by the electron
capture, can autoionize to the initial electronic state of a different
vibrational quantum number $v_{f}^{+}$ and then expel an electron to the
continuum. Vibrational excitation takes place when $v_{f}^{+} > v_{i}^{+}$,
while vibrational de-excitation occurs if $v_{f}^{+} < v_{i}^{+}$. The quantum
defect approach treats the processes represented by Eqs. (\ref{eq:direct}) and
(\ref{eq:indirect}) as multichannel reactive processes involving the
dissociation channels (accounting for the atom-atom scattering) and ionization
channels (accounting for the electron-molecular ion scattering). Each ionization
channel, for which the collision coordinate is the electron distance $r$ from
the molecular ion center, is defined by its threshold, a vibrational level $v^+$
of the molecular ion ground state and by the angular quantum number $l$ of the
incoming or outgoing electron. An ionisation channel is open if its
corresponding threshold is situated below the total energy of the system, and
closed otherwise. Each closed channels introduce into the calculations a
series of Rydberg states BeD$^*$ differing only by the principal quantum number of the
external electron~\cite{Schneider:94}. On the other hand, a dissociation channel,
having the internuclear distance $R$ as the collision coordinate, relies on an
electronically bound state BeD$^{**}$ whose potential energy in the asymptotic
limit is situated below the total energy of the system.

\begin{figure}[t]
\centering
\includegraphics[width=1.085\columnwidth]{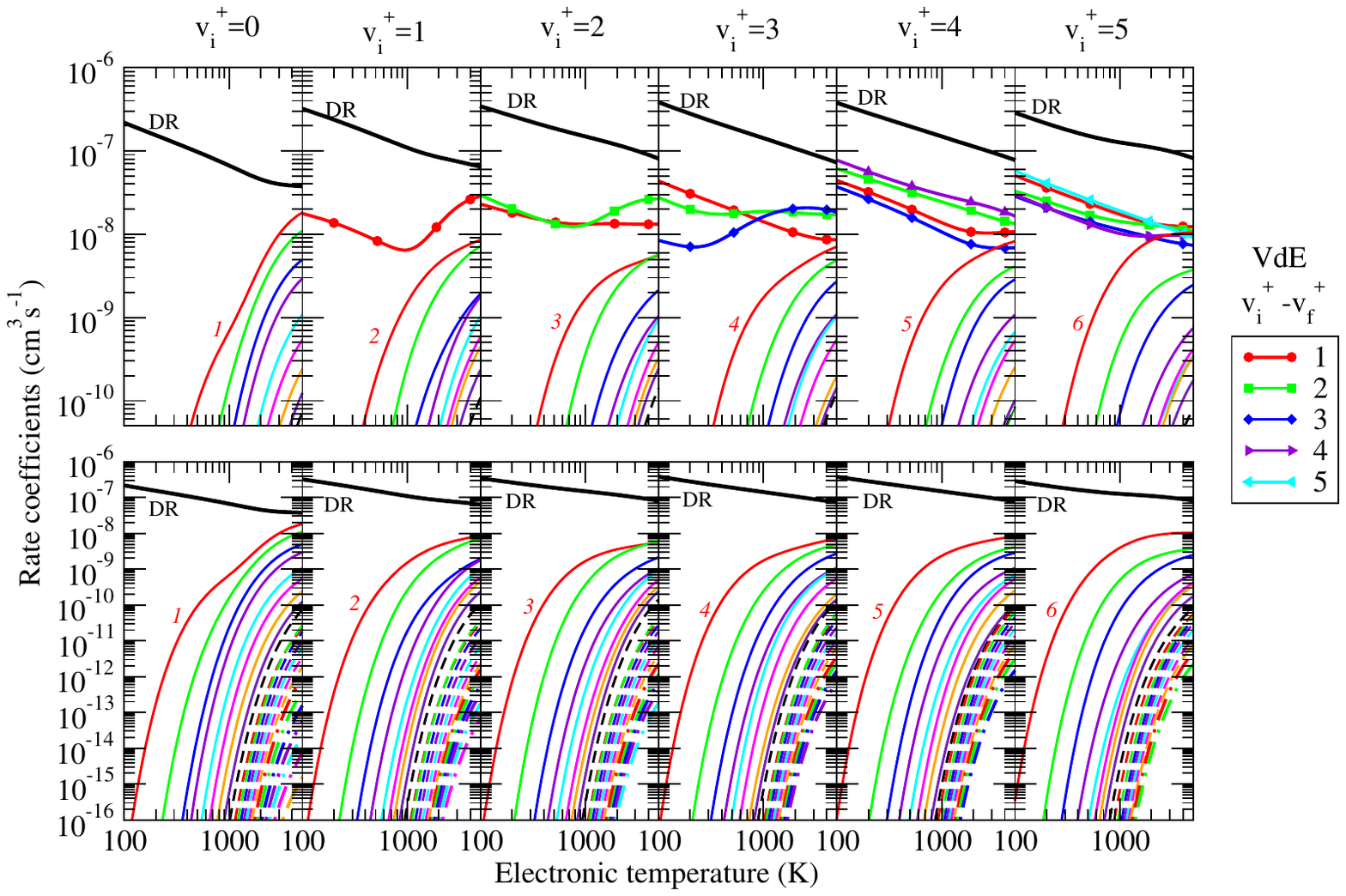}
\caption{Dissociative recombination (DR, black line), vibrational excitation (VE, thin lines) and vibrational de-excitation (VdE, symbols and thick lines) rate coefficients of {\rm BeD$^{+}$} in its electronic ground state for $v_{i}^{+}=0,\ldots5$. Upper panels: For each initial vibrational state of BeD$^+$, the final vibrational quantum numbers are labeled for de-excitation and for the first one vibrational excitation curve. The remaining unlabeled curves correspond to VE rate  coefficients of the ion in the successive increase (in the order) of the  vibrational quantum numbers. Lower panels: The same data without those of VdE process, the panels extending the range down to \rm{10$^{-16}$ cm$^3$/s}.\label{fig:5}}
\end{figure}
\begin{figure}[t]
\centering
\includegraphics[width=1.085\columnwidth]{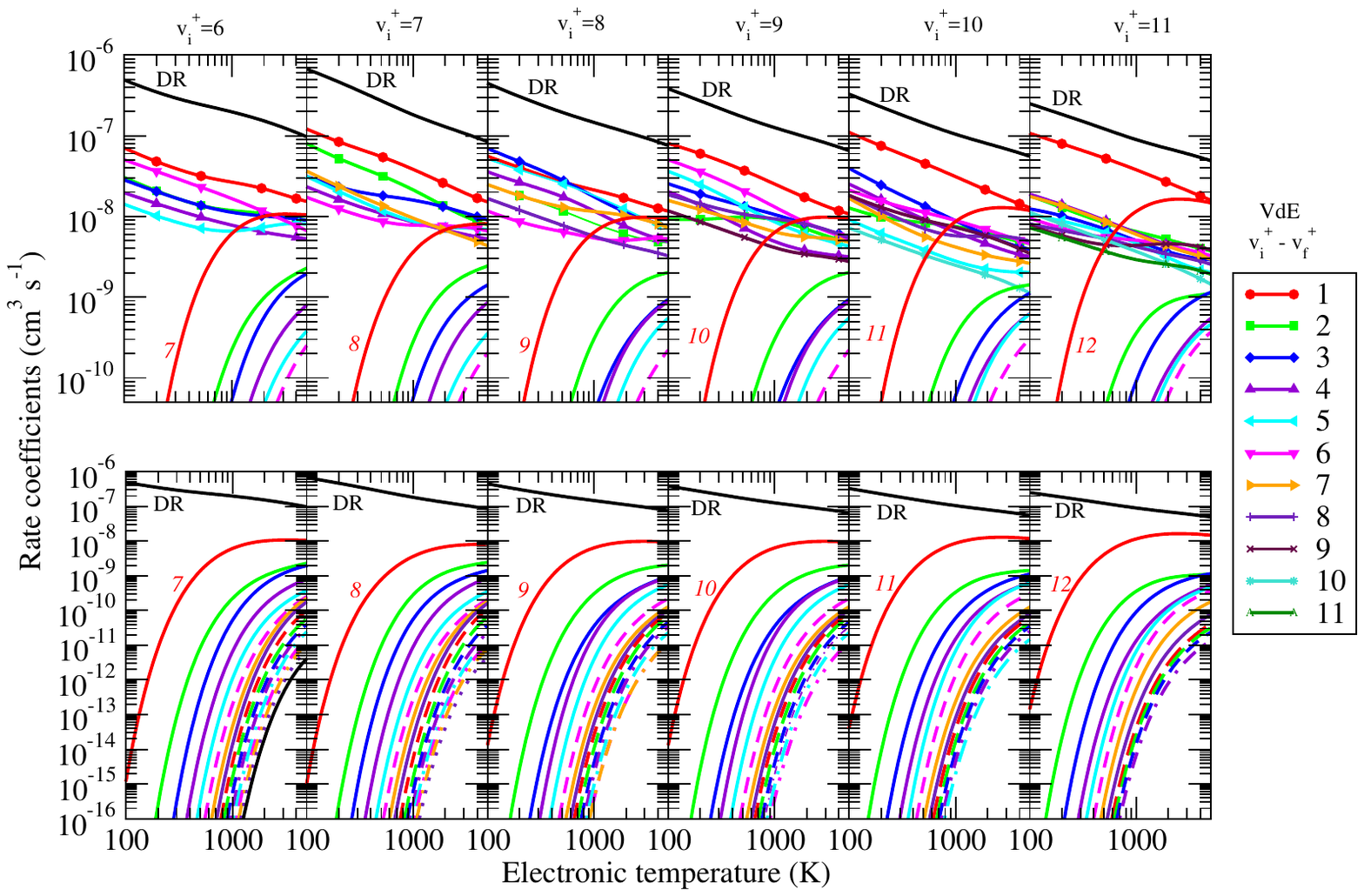}
\caption{Same as in Fig.~\ref{fig:5} for $v_{i}^{+}=6,\ldots 11$.\label{fig:6}}
%\end{center}
\end{figure}

Within the Born-Oppenheimer approximation, BeH$^{+}$ and BeD$^{+}$ ions having the same electronic structure, we used the
same set of potential energy curves, electronic couplings~\cite{roos:09} and quantum defects as in our previous work on BeH$^{+}$~\cite{niyonzima2016, niyonzima2013}, taking care to consider the reduced mass of BeD$^{+}$. Moreover, as the mathematical model is the same, we skip further details and we refer the reader to the previous article~\cite{niyonzima2016}.

Once the scattering matrix $S$ for the processes DR and VE/VdE are determined, the corresponding global cross sections, as a function of the incident electron kinetic energy $\varepsilon$, are obtained by summation over all relevant symmetries of the system and over the projection of the total electronic angular momentum on the nuclear axes $\Lambda$ of the resulting partial capture cross sections $\sigma$  into all the dissociative states $d_j$ of the same symmetry:
\begin{eqnarray}
\sigma _{diss \leftarrow v_{i}^{+}}(\varepsilon) &=& \frac{\pi}{4\varepsilon} \sum_{\Lambda,sym} \rho^{sym,\Lambda}\sum_{l,j}\left| S^\Lambda_{d_{j},lv_{i}^{+}}\right|^2\,,\label{eqDR}
\\
\sigma _{v_{f}^{+} \leftarrow v_{i}^{+}}(\varepsilon) &=& \frac{\pi}{4\varepsilon} \sum_{\Lambda,sym} \rho^{sym,\Lambda}\sum_{l,l'}\left| S^\Lambda_{l' v_{f}^{+},lv_{i}^{+}}-\delta_{l'l}\delta_{v_{i}^{+}v_{f}^{+}}\right|^2\,,\label{eqVE_VdE}
\end{eqnarray}
where $\rho^{sym,\Lambda}$ is the multiplicity ratio between the electronic states of BeD and the electronic states of BeD$^+$. In Eq.(\ref{eqVE_VdE}), the vibrational transition occurs \textit{via} the temporaly neutral BeD$^*$ (direct process) or  BeD$^{**}$ (indirect process) molecule electronic state. In Eq.(\ref{eqVE_VdE}), $l$ denotes the partial wave of the incident electron and $l'$ that of the outgoing electron. The total state multiplicities of the fragments (D and Be) are those of the BeD molecular state. Notice that the Eqs. (\ref{eqDR}) and (\ref{eqVE_VdE}) are written in atomic units.

In order to obtain the thermal rate coefficients, we have convoluted the global cross sections with the Maxwellian distribution function for velocities $v$ (related to incident energy of the electrons by $\varepsilon=\frac{1}{2}mv^{2}$) of the free electrons:
\begin{equation}\label{rate}
\alpha(T)=\frac{8\pi}{\sqrt m{(2\pi kT)}^{3/2}}\int_{0}^{+\infty}\,\sigma(\varepsilon)\,\varepsilon\,\exp(-\varepsilon/k_B T)\,d\varepsilon\,,
\end{equation}
where $\sigma(\varepsilon)$ is the cross sections given by (\ref{eqDR}) or (\ref{eqVE_VdE}), $k_B$ and $T$ being the Boltzman constant and the absolute temperature respectively.

\section{Results \label{res_disc}}
Using the available molecular data shown in Figure 1 of \cite{niyonzima2013} (for more details see as well \cite{roos:09,Strömholm1995}) - the  potential energy curves in a quasi-diabatic representation and electronic Rydberg-valence couplings 5\,\ensuremath{^{2}\Pi}, 5\,\ensuremath{^{2}\Sigma^{+}} and 1\,\ensuremath{^{2}\Delta} states - we have performed calculations corresponding to all vibrational levels (up to $v_i^+ =23$) of the ground electronic state of the ion. These 24 vibrational levels have been obtained in solving the radial Schroedinger equation by Numerov method, using the potential energy curve of BeH$^+$ electronic ground state from Ref. \cite{roos:09}. Table \ref{tab:viblev} shows the list of vibrational levels of BeD$^+$ and the values of $D_e$ and $D_0$. Notice that these values, as well as some of the potential energy curves and couplings for the neutral, are different from those of Ref. \cite{0741-3335-59-4-045008} since the molecular data have been obtained using different quantum chemical methods. In the following calculations, the energy of the electron is below to $2.7$ eV, this value being only slightly higher than the dissociation threshold $D_0$ of the ground electronic state of the ion.

\begin{figure}[t]
\centering
\includegraphics[width=1.085\columnwidth]{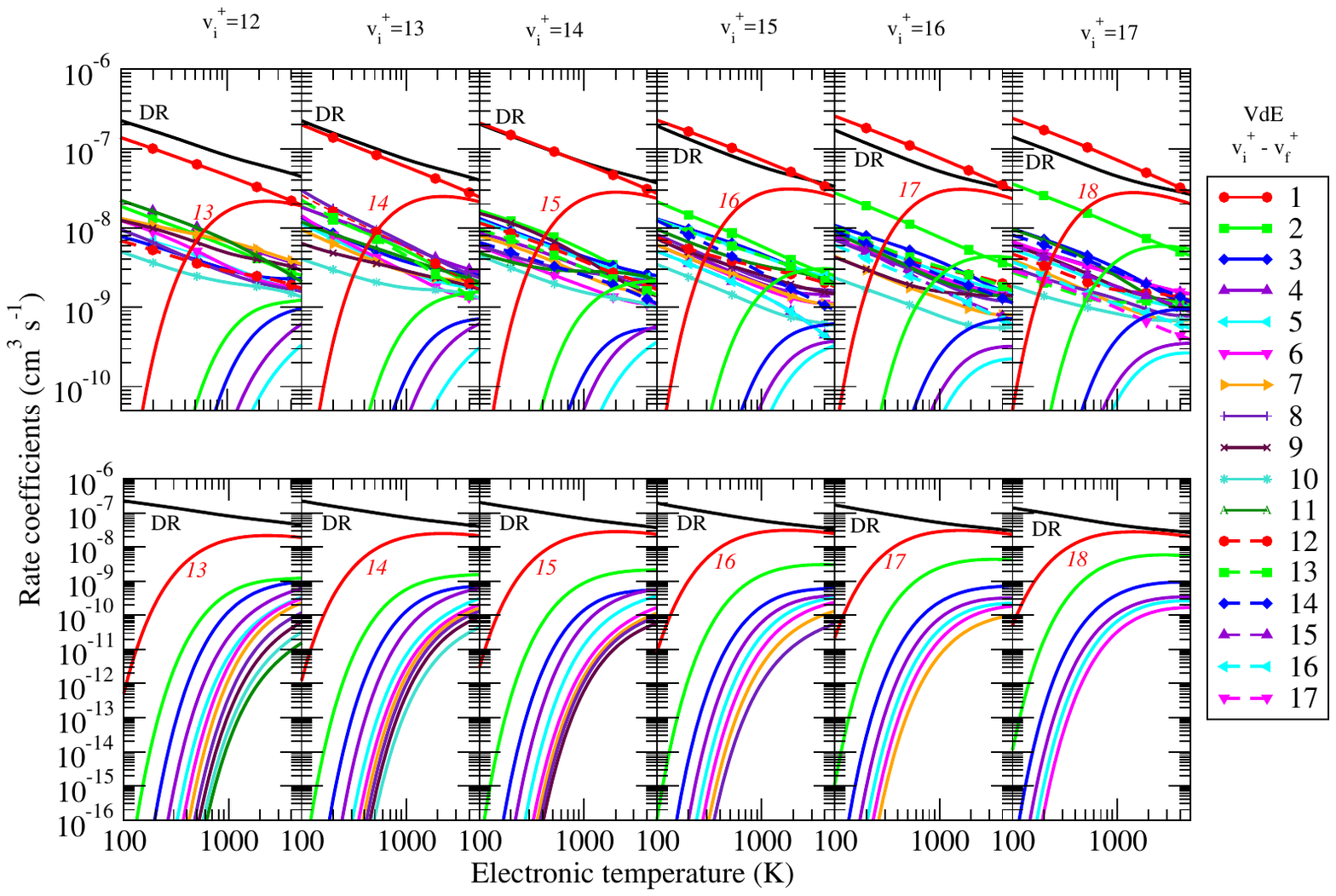}
\caption{Same as in Fig.~\ref{fig:5} for $v_{i}^{+}=12,\ldots 17$.\label{fig:7}}
\end{figure}
\begin{figure}[t]
\centering
\includegraphics[width=1.085\columnwidth]{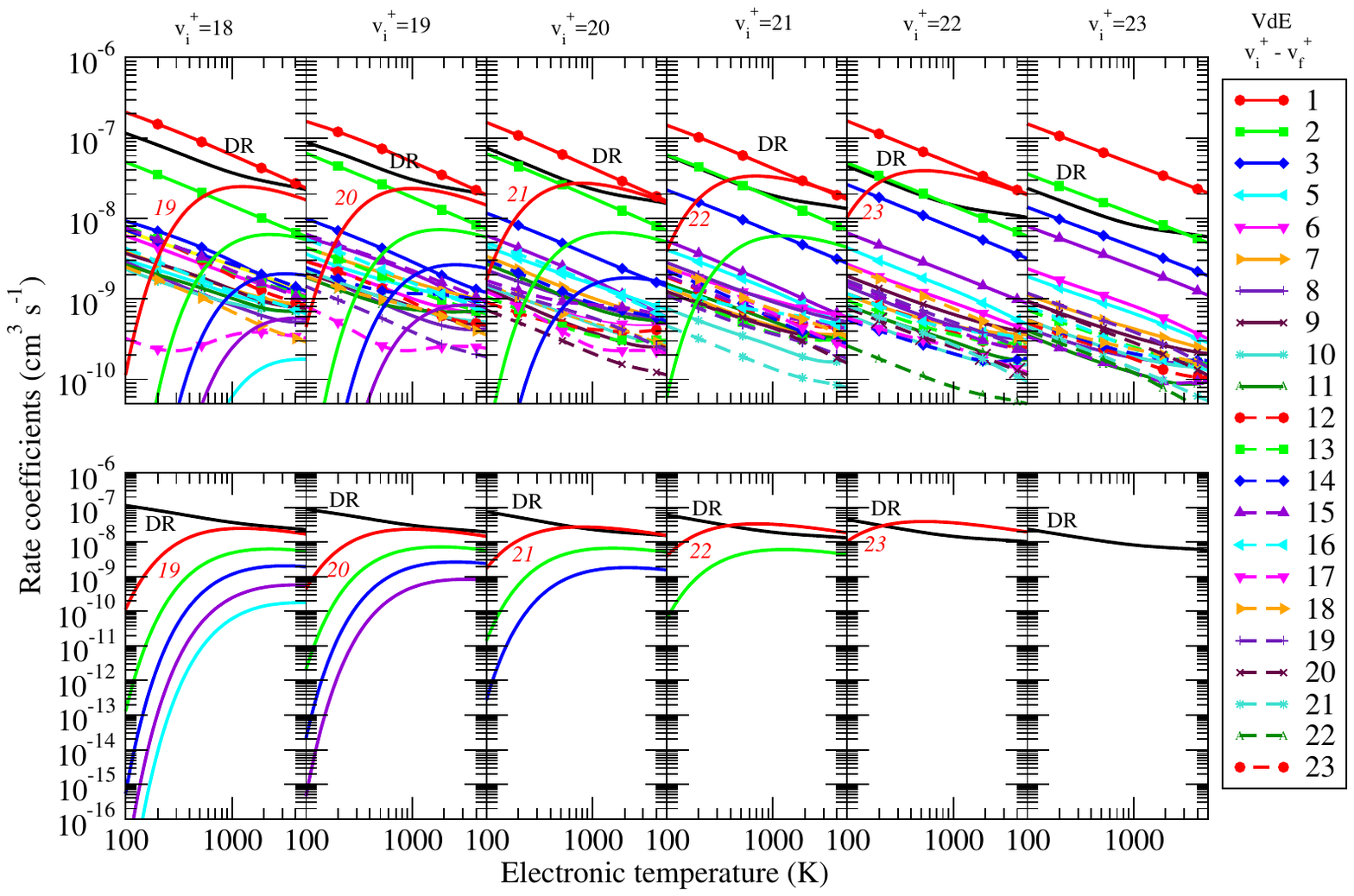}
\caption{Same as in Fig.~\ref{fig:5} for $v_{i}^{+}=18,\ldots 23$.\label{fig:8}}
%\end{center}
\end{figure}

\begin{table}
\centering
\begin{tabular}{cccc}
     \hline
$v^+$ & $\epsilon_{v^+}$(eV) & $v^+$ & $\epsilon_{v^+}$(eV) \\
     \hline\hline
0  & 0.000 &  12 & 1.920  \\
1  & 0.194 &  13 & 2.031  \\
2  & 0.380 &  14 & 2.135  \\
3  & 0.564 &  15 & 2.232  \\
4  & 0.742 &  16 & 2.321  \\
5  & 0.914 &  17 & 2.401  \\
6  & 1.079 &  18 & 2.473  \\
7  & 1.238 &  19 & 2.535  \\
8  & 1.391 &  20 & 2.585  \\
9  & 1.537 &  21 & 2.623  \\
10 & 1.674 &  22 & 2.655  \\
11 & 1.802 &  23 & 2.678  \\
     \hline
\end{tabular}
\caption{BeD$^+$ vibrational levels referred to $v^+=0$. The values of dissociating energies are $D_e=2.794$ eV and $D_0=2.682$ eV. \label{tab:viblev}}
\end{table}

\begin{figure*}[t]
\centering
\includegraphics[width=.75\textwidth]{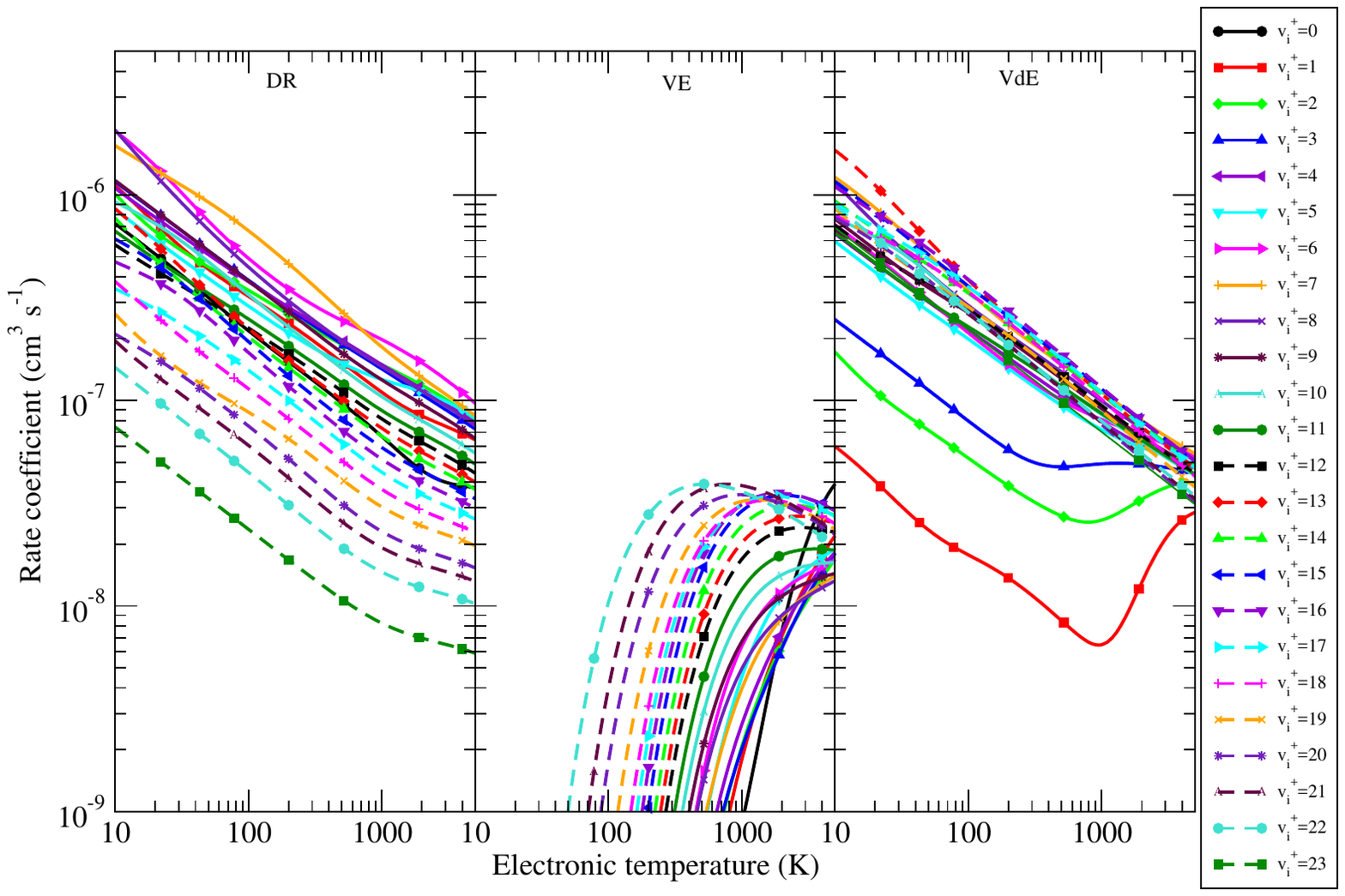}
\caption{Dissociative recombination (DR), vibrational excitation (VE) and vibrational de-excitation (VdE)
rate coefficients for various initial vibrational states indicated in the legend,
those of VE and VdE processes being obtained by sum over all the final states. \label{fig:9}}
%\end{center}
\end{figure*}

Figures \ref{fig:5}--\ref{fig:8} give the whole ensemble of rate coefficients
available for the state-to-state kinetics of BeD$^+$. They illustrate the fact that DR dominates for
$v_{i}^{+}=0-13$ levels at low electron temperature,
while the VdE becomes more important than the other processes for initial
vibrational states $v_{i}^{+}>13$. Figure \ref{fig:9} provides a comparison
between the DR rate coefficients and the global vibrational transitions rate
coefficients - \emph{i.e.} coming from the
sum over all the possible final levels. The excitation process competes with DR and VdE above $1000$~K only.

The rate coefficients shown in Figs.~\ref{fig:5}-\ref{fig:8} have been fitted by a generalized Arrhenius-type formulas in order to be ready for use in  codes for kinetics modelling. The calculated DR rate coefficients of BeD$^+$ in each of its first 24 vibrational states ($v_i^+=0\ldots23$) have been interpolated under the mathematical form:
\begin{equation}
k_{v_i^+}^{DR}(T_e) = A_{v_i^+} \, T_e^{\alpha_{v_i^+}} \, \exp\left[-\sum_{j=1}^{7} \frac{B_{v_i^+}(j)}{j\,\,T_e^j} \right]\,,\label{eqn:BeD_DR_Interpolation}
\end{equation}
over the electron temperature range $100\le T_e\le 5000$~K. The parameters
$A_{v_i^+}$, $\alpha_{v_i^+}$ and $B_{v_i^+}(j)$ are listed in the Table \ref{tab:BeD_DR_Interpolation}. The calculated VE and VdE rate coefficients of BeD$^+$ have been interpolated under the form:
\begin{equation}
k_{v_i^+\to v_f^+}^{V(d)E}(T_e) = A_{v_i^+\to v_f^+} \, T_e^{\alpha_{v_i^+\to v_f^+}} \, \exp\left[-\sum_{j=1}^{7}\frac{B_{v_i^+\to v_f^+}(j)}{j\,\,T_e^j}\right]\,,\label{eqn:BeD_VE_VdE_Interpolation}
\end{equation}
over the electron temperature range $300\le T_e\le 5000$~K.

The parameters $A_{v_i^+\to v_f^+}$, $\alpha_{v_i^+\to v_f^+}$ and $B_{v_i^+\to v_f^+}(j)$ are listed in the Table \ref{tab:BeD_VE_VdE_Interpolation0-1} for the single quantum VE, \emph{i.e.} $v_f^+=v_i^+ + 1$. The fitted values obtained by Eqs.~(\ref{eqn:BeD_DR_Interpolation}) and (\ref{eqn:BeD_VE_VdE_Interpolation}) depart from the calculated values only of a few percent. The full set of coefficients for DR, VE and VdE are given in the supplementary material of the present article.

\begin{table*}[t]
\renewcommand{\arraystretch}{1.00} % enlarge line spacing
\centering
	\caption{\label{tab:BeD_DR_Interpolation}List of the parameters used in Eq.~(\ref{eqn:BeD_DR_Interpolation}) for the DR rate coefficients of BeD$^+$.}
  \footnotesize
%  \scriptsize
%	\rotatebox{90}{
    \begin{tabular}{cccccccccc}
		\hline
    $v_i^+$ &  $A_{v_i^+}$  & $\alpha_{v_i^+}$                  &  $B_{v_i^+}(1)$  & $B_{v_i^+}(2)$   & $B_{v_i^+}(3)$  & $B_{v_i^+}(4)$  & $B_{v_i^+}(5)$   & $B_{v_i^+}(6)$  & $B_{v_i^+}(7)$ \\
 & (cm$^3\cdot$s$^{-1}\cdot$K$^{-\alpha}$) & & ($\times10^{3}$ K) &($\times10^{6}$ K$^2$) &($\times10^{9}$ K$^3$) &($\times10^{11}$ K$^4$) &($\times10^{14}$ K$^5$) &($\times10^{15}$ K$^6$) &($\times10^{17}$ K$^7$) \\ 
		\hline
 0    & $    0.8974\times10^{-09 }$ & $    0.3841$ & $   -2.563$ & $    2.470$ & $   -1.13$ & $    2.766$ & $   -0.3713$ & $    2.571$ & $   -0.7179$    \\
 1    & $    0.6299\times10^{-06 }$ & $   -0.2688$ & $   -0.0219$ & $   -0.3068$ & $    0.2394$ & $   -0.7544$ & $    0.1177$ & $   -0.9028$ & $    0.2710$    \\
 2    & $    0.1395\times10^{-04 }$ & $   -0.5889$ & $    0.7304$ & $   -0.7024$ & $    0.3170$ & $   -0.7705$ & $    0.1032$ & $   -0.7156$ & $    0.2002$    \\
 3    & $    0.3307\times10^{-05 }$ & $   -0.4462$ & $    0.09445$ & $   -0.08147$ & $    0.03345$ & $   -0.07925$ & $    0.01074$ & $   -0.07665$ & $    0.02216$    \\
 4    & $    0.4349\times10^{-05 }$ & $   -0.4673$ & $    0.2376$ & $   -0.2605$ & $    0.1310$ & $   -0.3448$ & $    0.04893$ & $   -0.3541$ & $    0.1023$    \\
 5    & $    0.2623\times10^{-04 }$ & $   -0.6473$ & $    1.421$ & $   -1.540$ & $    0.7367$ & $   -1.848$ & $    0.2520$ & $   -1.766$ & $    0.4973$    \\
 6    & $    0.3822\times10^{-04 }$ & $   -0.6855$ & $    0.7458$ & $   -0.5492$ & $    0.1993$ & $   -0.4139$ & $    0.04962$ & $   -0.3181$ & $    0.08418$    \\
 7    & $    0.6649\times10^{-05 }$ & $   -0.5080$ & $    0.2471$ & $   -0.4732$ & $    0.2797$ & $   -0.7808$ & $    0.1134$ & $   -0.8287$ & $    0.2402$    \\
 8    & $    0.7614\times10^{-05 }$ & $   -0.5318$ & $    0.4697$ & $   -0.5195$ & $    0.2456$ & $   -0.6106$ & $    0.08275$ & $   -0.5776$ & $    0.1623$    \\
 9    & $    0.6493\times10^{-05 }$ & $   -0.5295$ & $    0.5057$ & $   -0.6108$ & $    0.3022$ & $   -0.7682$ & $    0.1054$ & $   -0.7416$ & $    0.2094$    \\
10    & $    0.3819\times10^{-05 }$ & $   -0.4877$ & $    0.4455$ & $   -0.6156$ & $    0.3269$ & $   -0.8697$ & $    0.1230$ & $   -0.8836$ & $    0.2532$    \\
11    & $    0.2883\times10^{-05 }$ & $   -0.4686$ & $    0.4699$ & $   -0.6798$ & $    0.3766$ & $   -1.028$ & $    0.1479$ & $   -1.074$ & $    0.3105$    \\
12    & $    0.2315\times10^{-05 }$ & $   -0.4563$ & $    0.4047$ & $   -0.6130$ & $    0.3466$ & $   -0.9566$ & $    0.1385$ & $   -1.011$ & $    0.2931$    \\
13    & $    0.1442\times10^{-05 }$ & $   -0.4155$ & $    0.2709$ & $   -0.4825$ & $    0.2839$ & $   -0.8004$ & $    0.1174$ & $   -0.8643$ & $    0.2521$    \\
14    & $    0.9694\times10^{-06 }$ & $   -0.3803$ & $    0.1729$ & $   -0.3907$ & $    0.2413$ & $   -0.6932$ & $    0.1026$ & $   -0.7602$ & $    0.2226$    \\
15    & $    0.7270\times10^{-06 }$ & $   -0.3590$ & $    0.1667$ & $   -0.4139$ & $    0.2543$ & $   -0.7253$ & $    0.1068$ & $   -0.7881$ & $    0.2301$    \\
16    & $    0.4768\times10^{-06 }$ & $   -0.3232$ & $    0.1124$ & $   -0.3870$ & $    0.2458$ & $   -0.7065$ & $    0.1042$ & $   -0.7692$ & $    0.2245$    \\
17    & $    0.3339\times10^{-06 }$ & $   -0.2965$ & $    0.08587$ & $   -0.3909$ & $    0.2567$ & $   -0.7453$ & $    0.1104$ & $   -0.8167$ & $    0.2387$    \\
18    & $    0.2383\times10^{-06 }$ & $   -0.2735$ & $    0.1221$ & $   -0.4597$ & $    0.2953$ & $   -0.8485$ & $    0.1250$ & $   -0.9205$ & $    0.2683$    \\
19    & $    0.1868\times10^{-06 }$ & $   -0.2609$ & $    0.1973$ & $   -0.5555$ & $    0.3423$ & $   -0.9644$ & $    0.1404$ & $   -1.025$ & $    0.2974$    \\
20    & $    0.1332\times10^{-06 }$ & $   -0.2495$ & $    0.2343$ & $   -0.6118$ & $    0.3644$ & $   -1.009$ & $    0.1454$ & $   -1.056$ & $    0.3048$    \\
21    & $    0.1114\times10^{-06 }$ & $   -0.2452$ & $    0.2826$ & $   -0.6493$ & $    0.3776$ & $   -1.035$ & $    0.1486$ & $   -1.076$ & $    0.3102$    \\
22    & $    0.8079\times10^{-07 }$ & $   -0.2366$ & $    0.2937$ & $   -0.6480$ & $    0.3747$ & $   -1.027$ & $    0.1475$ & $   -1.069$ & $    0.3084$    \\
23    & $    0.4801\times10^{-07 }$ & $   -0.2398$ & $    0.3455$ & $   -0.7133$ & $    0.4152$ & $   -1.148$ & $    0.1659$ & $   -1.208$ & $    0.3498$    \\

		\hline
    \end{tabular}
    %}%
    \normalsize
\end{table*}%

\begin{table*}[t]
\renewcommand{\arraystretch}{1.00} % enlarge line spacing
\centering
	\caption{\label{tab:BeD_VE_VdE_Interpolation0-1}List of the parameters used in Eq.~(\ref{eqn:BeD_VE_VdE_Interpolation}) for the monoquantic VE, $v_i^+ \to v_f^+=v_i^++1$,  rate coefficients of BeD$^+$.}
\footnotesize
%  \scriptsize
%	\rotatebox{90}{
    \begin{tabular}{cccccccccc}
		\hline
    $v_i^+$ &  $A_{v_i^+\to v_f^+}$ & $\alpha_{v_i^+\to v_f^+}$         & $B_{v_i^+\to v_f^+}(1)$ & $B_{v_i^+\to v_f^+}(2)$ & $B_{v_i^+\to v_f^+}(3)$  & $B_{v_i^+\to v_f^+}(4)$ & $B_{v_i^+\to v_f^+}(5)$ & $B_{v_i^+\to v_f^+}(6)$ & $B_{v_i^+\to v_f^+}(7)$ \\
 & (cm$^3\cdot$s$^{-1}\cdot$K$^{-\alpha}$) & & ($\times10^{4}$ K) &($\times10^{7}$ K$^2$) &($\times10^{10}$ K$^3$) &($\times10^{13}$ K$^4$) &($\times10^{15}$ K$^5$) &($\times10^{18}$ K$^6$) &($\times10^{19}$ K$^7$) \\ 
		\hline
 0    & $    0.232\times10^{+02 }$ & $   -2.10$ & $    1.68$ & $   -1.56$ & $   -0.611$ & $    1.73$ & $   -10.1$ & $    2.51$ & $   -23.0$    \\
 1    & $    0.395\times10^{-09 }$ & $    0.361$ & $   -0.0592$ & $    0.698$ & $   -0.787$ & $    0.442$ & $   -1.28$ & $    0.176$ & $   -0.830$    \\
 2    & $    0.156\times10^{-11 }$ & $    0.887$ & $   -0.423$ & $    1.26$ & $   -1.35$ & $    0.833$ & $   -2.94$ & $    0.551$ & $   -4.24$    \\
 3    & $    0.240\times10^{-06 }$ & $   -0.294$ & $    0.638$ & $   -1.58$ & $    2.16$ & $   -1.50$ & $    5.67$ & $   -1.10$ & $    8.67$    \\
 4    & $    0.109\times10^{-07 }$ & $    0.0162$ & $    0.249$ & $   -0.457$ & $    0.739$ & $   -0.555$ & $    2.18$ & $   -0.439$ & $    3.53$    \\
 5    & $    0.432\times10^{-08 }$ & $    0.0776$ & $   -0.205$ & $    1.13$ & $   -1.44$ & $    0.960$ & $   -3.51$ & $    0.669$ & $   -5.19$    \\
 6    & $    0.517\times10^{-08 }$ & $    0.0507$ & $   -0.211$ & $    0.892$ & $   -1.0$ & $    0.639$ & $   -2.29$ & $    0.435$ & $   -3.36$    \\
 7    & $    0.753\times10^{-08 }$ & $   -0.00187$ & $   -0.107$ & $    0.714$ & $   -0.824$ & $    0.509$ & $   -1.75$ & $    0.318$ & $   -2.36$    \\
 8    & $    0.128\times10^{-07 }$ & $   -0.0677$ & $   -0.0679$ & $    0.444$ & $   -0.448$ & $    0.262$ & $   -0.889$ & $    0.160$ & $   -1.19$    \\
 9    & $    0.220\times10^{-06 }$ & $   -0.353$ & $    0.0627$ & $    0.160$ & $   -0.141$ & $    0.0746$ & $   -0.234$ & $    0.0402$ & $   -0.290$    \\
10    & $    0.143\times10^{-05 }$ & $   -0.530$ & $    0.138$ & $    0.00943$ & $    0.0135$ & $   -0.0209$ & $    0.107$ & $   -0.0249$ & $    0.220$    \\
11    & $    0.380\times10^{-05 }$ & $   -0.613$ & $    0.172$ & $   -0.0627$ & $    0.0732$ & $   -0.0503$ & $    0.195$ & $   -0.0394$ & $    0.321$    \\
12    & $    0.850\times10^{-05 }$ & $   -0.673$ & $    0.196$ & $   -0.135$ & $    0.151$ & $   -0.0976$ & $    0.358$ & $   -0.0693$ & $    0.547$    \\
13    & $    0.144\times10^{-04 }$ & $   -0.718$ & $    0.216$ & $   -0.203$ & $    0.230$ & $   -0.148$ & $    0.542$ & $   -0.104$ & $    0.824$    \\
14    & $    0.129\times10^{-04 }$ & $   -0.698$ & $    0.192$ & $   -0.167$ & $    0.184$ & $   -0.114$ & $    0.404$ & $   -0.0755$ & $    0.578$    \\
15    & $    0.160\times10^{-04 }$ & $   -0.719$ & $    0.193$ & $   -0.204$ & $    0.237$ & $   -0.153$ & $    0.556$ & $   -0.106$ & $    0.827$    \\
16    & $    0.153\times10^{-04 }$ & $   -0.723$ & $    0.181$ & $   -0.207$ & $    0.247$ & $   -0.161$ & $    0.591$ & $   -0.113$ & $    0.893$    \\
17    & $    0.103\times10^{-04 }$ & $   -0.700$ & $    0.153$ & $   -0.162$ & $    0.189$ & $   -0.121$ & $    0.434$ & $   -0.0817$ & $    0.630$    \\
18    & $    0.784\times10^{-05 }$ & $   -0.691$ & $    0.137$ & $   -0.150$ & $    0.176$ & $   -0.113$ & $    0.405$ & $   -0.0765$ & $    0.589$    \\
19    & $    0.401\times10^{-05 }$ & $   -0.638$ & $    0.0942$ & $   -0.0839$ & $    0.0939$ & $   -0.0576$ & $    0.199$ & $   -0.0365$ & $    0.276$    \\
20    & $    0.323\times10^{-05 }$ & $   -0.609$ & $    0.0779$ & $   -0.0786$ & $    0.0904$ & $   -0.0574$ & $    0.205$ & $   -0.0387$ & $    0.298$    \\
21    & $    0.273\times10^{-05 }$ & $   -0.573$ & $    0.0600$ & $   -0.0489$ & $    0.0528$ & $   -0.0319$ & $    0.109$ & $   -0.0200$ & $    0.150$    \\
22    & $    0.221\times10^{-05 }$ & $   -0.545$ & $    0.0446$ & $   -0.0367$ & $    0.0403$ & $   -0.025$ & $    0.0878$ & $   -0.0162$ & $    0.123$    \\
		\hline
    \end{tabular}
    %}%
    \normalsize
\end{table*}%

\section{Conclusions \label{sec:concl}}

The present paper provides complete set of vibrational resolved rate coefficients for BeD$^{+}$ cation reactive
collisions with electrons below to the ion dissociation threshold. In particular, the competition between the vibrational transitions and dissociative recombination processes are illustrated quantitatively. Arrhenius-type formulas are used for fitting the rate coefficients as function of the electron temperature. The rate coefficients are strongly dependent on the initial vibrational level of the molecular ion.

These data are relevant for the modeling of the edge of the fusion plasma. The higher energy region, where the dissociative excitation process \cite{0741-3335-59-4-045008, Kalyan2013} competes the dissociative recombination and the vibrational transitions, as well as similar calculations on BeT$^+$, are object of ongoing work.

\section*{Acknowledgments}

SN is grateful to the project ``Projet 1 du Programme CUI: Universit\'e du Burundi -VLIR'' for the research stay in KULeuven during which the major part of the work has been carried out. VL, JZM, AB and IFS acknowledge the French LabEx EMC$^3$, \emph{via} the projects PicoLIBS (ANR-12-BS05-0011-01) and EMoPlaF, the BIOENGINE project and the VIRIDIS-CO2 project (sponsored by the European
fund FEDER and the French CPER), the F\'ed\'eration de Recherche Fusion par Confinement Magn\'etique - ITER and the European COST Program CM1401 (Our Astrochemical History). JZM is grateful for financial support from Labex SEAM and IDEX-USPC. We are also grateful to the French 
\section*{Data availability}
Upon a reasonable request, the data supporting this article will be provided by the corresponding author.

\end{document}